\pdfoutput=1
\documentclass[5p,times]{elsarticle}
\journal{Physics Letters B}

\usepackage[colorlinks=true, pdfstartview=FitV, linkcolor=red, citecolor=blue,
urlcolor=blue]{hyperref}
\usepackage{graphicx}
\usepackage{amsmath,amssymb}
\usepackage{color}

\newcommand{\Lag}{\mathcal{L}}
\newcommand{\calS}{\mathcal{S}}

\newcommand{\tr}{\text{tr}}
\newcommand{\Tr}{\text{Tr}}
\newcommand{\bOmega}{\boldsymbol{\Omega}}
\newcommand{\bB}{\boldsymbol{B}}

\newcommand{\bj}{\boldsymbol{j}}

\newcommand{\bz}{\boldsymbol{z}}

\newcommand{\GeV}{\;\text{GeV}}

\usepackage[normalem]{ulem}

\newcommand\rout{\bgroup \color{red} \ULdepth=-.5ex \ULset}
\newcommand\bout{\bgroup \color{blue} \ULdepth=-.5ex \ULset}

\begin{document}

\title{Boundary effects and gapped dispersion in rotating fermionic matter}

\author{Shu Ebihara}
\author{Kenji Fukushima}
\author{Kazuya Mameda}
\address{Department of Physics, The University of Tokyo,
         7-3-1 Hongo, Bunkyo-ku, Tokyo 113-0033, Japan}
\begin{abstract}
  We discuss the importance of boundary effects on fermionic matter in
  a rotating frame.  By explicit calculations at zero temperature we
  show that the scalar condensate of fermion and anti-fermion cannot
  be modified by the rotation once the boundary condition is properly
  implemented.  The situation is qualitatively changed at finite
  temperature and/or in the presence of a sufficiently strong magnetic
  field that supersedes the boundary effects.  Therefore, to establish
  an interpretation of the rotation as an effective chemical
  potential, it is crucial to consider further environmental effects
  such as the finite temperature and magnetic field.
\end{abstract}
\maketitle

\section{Introduction}

Examples of relativistic fermionic systems with substantial angular
velocity are found in matter in extreme environments such as the cores
of spinning compact stellar objects, the merger of binary stars, the
heavy-ion collision, etc.  In early days pioneering
works~\cite{Vilenkin:1978hb,Vilenkin:1979ui,Vilenkin:1980zv} were
motivated by astrophysical applications, and nowadays, the theoretical
interest in relativistic rotating systems is being revived inspired by
non-central collisions of heavy ions where a global spin polarization
could be measurable~\cite{Liang:2004ph,Huang:2011ru,Huang:2015oca,
Becattini:20082452,Becattini:201332,Becattini:2015ko,Jiang:2016woz,%
Aristova:2016wxe,Deng:2016gyh}.
Although it is technically difficult to design a directly rotating
experiment, a material of Dirac/Weyl semimetal under circular
polarized electromagnetic fields (as considered in
Ref.~\cite{Ebihara:2015aca} for instance) can be also understood in
the same way as a rotating system, which is evident with an
appropriate Floquet transformation~\cite{Bukov:2014}.  From an
intuitive analogy between the angular momentum and the magnetic field
and a further formal similarity with Landau quantization-like
effects~\cite{Wilkin:2000zz,Fetter:2009zz,Mameda:2015ria},  it has
been anticipated that the rotation should induce as exotic phenomena
as the magnetic field would do;  one typical example is the
topological current induced by vorticity (local
rotation)~\cite{Son:2004tq,Son:2009tf}, which is called the chiral
vortical effect and is analogous to the chiral magnetic
effect~\cite{Kharzeev:2007jp,Fukushima:2008xe}.  It is also pointed
out that the chiral vortical effect has origins not only from the
gauge sector but also from the gravitational (mixed)
anomaly~\cite{Landsteiner:2011cp,Basar:2013qia}.  These currents may
affect some of the condensates and the ground state structure
too~\cite{Fukushima:2013zga}.

For the magnetic effects on the ground state structure, the best known
and understood is an inevitable formation of the scalar condensate
leading to spontaneous breaking of chiral symmetry, which is referred
to as the magnetic catalysis
~\cite{Klimenko:1991he,Klimenko:1992ch,Gusynin:1994re,Gusynin:1995nb}.  
It is thus a quite natural idea to expect a rotational counterpart
that affects some of the condensates.  The question we are going to
address is how the scalar condensate of fermion and anti-fermion
(which will be called the ``chiral condensate'' hereafter) should be
influenced by the rotation.

In a preceding work by two of the present authors and two 
colleagues~\cite{Chen:2015hfc}, it has been demonstrated by explicit
calculations that the rotation generates a term that can be
interpreted as an effective chemical potential and such a masqueraded
density manifests itself in a form of the finite-density inverse
magnetic catalysis~\cite{Ebert:1999ht,Preis:2010cq} under a strong
magnetic field.  This finiteness of density is a genuine physical
consequence beyond formal similarity and it arises from the quantum
anomaly as argued in Ref.~\cite{Hattori:2016njk} (that is also closely
related to the chiral pumping effect~\cite{Ebihara:2015aca}) 
\textit{if} a strong magnetic field is imposed.

Recently, a speculative scenario has been proposed about phase
transitions caused solely by angular velocity~\cite{Jiang:2016wvv}.
Because the chiral condensate melts at sufficiently high density, it
is likely that there is a critical value of the angular velocity,
$\Omega_c$, above which the chiral condensate is vanishing.  According
to the estimate in Ref.~\cite{Jiang:2016wvv} a first-order phase
transition takes place at $\Omega_c\simeq 0.65\GeV$ for rotating quark
matter sitting at $r=0.1\GeV^{-1}$ described by an effective model
with four-fermion interaction.  The purpose of this paper is to
investigate the finite radius effects of rotating fermionic matter.
In Ref.~\cite{Jiang:2016wvv} the finite size effect has been partially
taken into account in the local density approximation, but we will
point out that not only the local information at $r$ but also the bulk
boundary effects at $R$ would become as important.

Before looking into calculation details, let us give a hand-waving
argument:
The effective chemical potential in a rotating frame  with 
$\bOmega=\Omega \hat z$ is
characterized by $|\Omega j|$ where $j$ is the $z$-component of the
total angular momentum.  Therefore, fermionic modes with the energy
lower than $|\Omega j|$ are Pauli blocked.  If $\Omega$ goes larger,
more modes are blocked and eventually formation of condensation
could be hindered, which is an account for possible phase
transitions \textit{unless} the boundary effect is properly
implemented.  In finite size systems, the infrared (IR) cutoff is
introduced and the momenta should be discrete.  Hence, the fermion
energy dispersion should have a gap of order of $\sim R^{-1}$.
Because the wave-function with larger $j$ tends to have a more
spreading configuration profile due to the centrifugal force, it costs
a more energy in effect to confine the system in a cylinder, and
accordingly the energy gap should also increase as $\sim R^{-1}|j|$.
It would be then a delicate quantitative competition which of
$|\Omega j|$ and $R^{-1}|j|$ can be larger.  Our explicit calculations
(at zero temperature) will show that the energy gap
$\sim R^{-1}|j|$ is \textit{always} larger than the effective chemical
potential $|\Omega j|$, so that no mode is actually Pauli blocked.
This means that the chiral condensate cannot be modified at all so
long as the temperature is smaller than the effective chemical
potential.

We append two brief comments about the above intuitive argument.
First, we assume the quasiparticle approximation, which contains only 
the leading order contribution in the systematic expansion with respect 
to internal degrees of freedom (such as the number of the color of quarks).
Since the contributions from the fermionic paired states (e.g. mesons) are 
the next higher order, the aforementioned argument is correct within the 
four fermion interaction model, which consists of the leading order terms.
Besides even including bosonic states our argument should not to change
because in the bosonic case the rotational energy shift cannot exceed 
the boundary gap, as discussed in Refs.~\cite{Vilenkin:1980zv,%
Davies:1996ks}.

Second is about the difference of the rotational effects 
on fermions and antifermions.  While the authentic chemical potential
affects fermions and antifermions oppositely, the rotational energy
shift influences fermions and antifermions similarly.  Hence, both
fermions and antifermions (and thus both $j>0$ and $j<0$ states)
contribute to dynamics in rotating systems unlike the finite-density
case at zero temperature.  For example, if a fermion with angular momentum 
$j$ forms the chiral condensate, the partner antifermion should have $-j$ 
because the chiral condensate is a scalar paired state, with zero total 
angular momentum~\cite{Jiang:2016wvv}.  Therefore, in the Pauli blocking 
argument, as long as the authentic chemical potential is absent, fermions 
and antifermions and thus $j>0$ and $j<0$ states equally make a 
contribution and only the modulus of $\Omega j$ matters.

Our results imply that the phase transition scenario needs judicious
refinements in the low-temperature region.  At finite temperatures the
situation could be qualitatively changed, because there is no strict
Pauli blocking, and moreover the anomalous effects are turned on.  In
the end we will briefly mention on non-trivial interplay between the
rotation and the finite temperature and magnetic field.

\section{Reviewing the Dirac equation in a rotating frame}

We explain our notation by making a quick summary of basic formulas
for Dirac fermions in a rotating frame.  The free Dirac equation in
curved spacetime reads~\cite{birrell1984quantum},
\begin{equation}
  \bigl[ i\gamma^\mu (\partial_\mu + \Gamma_\mu) - m \bigr]\psi = 0\;,
\label{eq:Diraceq}
\end{equation}
where the covariant derivatives associated with finite rotation are
specified as $\Gamma_\mu=-\frac{i}{4}\omega_{\mu ij}\sigma^{ij}$ with
the Dirac spin matrices $\sigma^{ij}=\frac{i}{2}[\gamma^i,\gamma^j]$.
The spin connection is given by
$\omega_{\mu ij} = g_{\alpha\beta}e^\alpha_i
(\partial_\mu e^\beta_j + \Gamma^\beta_{\mu\nu}e^\nu_j)$
in terms of the metric and the vierbein, where Greek and Latin letters
represent coordinate $(\mu=t,x,y,z)$ and tangent $(i=0,1,2,3)$ space,
respectively.  In a rotating frame with the angular frequency vector,
${\boldsymbol \Omega} = \Omega\hat{\bz}$, we can write down the 
explicit form of the metric as
\begin{equation}
  g_{\mu\nu} = \begin{pmatrix}
    1-(x^2+y^2)\Omega^2  & y\Omega  & -x\Omega  &  0 \\
    y\Omega  &  -1  &  0  &  0 \\
    -x\Omega  &  0  &  -1  &  0 \\
    0  &  0  &  0  &  -1 \\
  \end{pmatrix}\;.
  \label{eq:metric}
\end{equation}
The corresponding vierbein is not unique and for convenience we shall
choose them as
\begin{equation}
  e^t_0 = e^x_1 = e_2^y = e^z_3 = 1, \qquad
  e^x_0 = y\Omega, \qquad
  e^y_0 = -x\Omega\;,
  \label{eq:vierbein}
\end{equation}
and zero for the other components.  We can simplify the Dirac matrix
structure of Eq.~\eqref{eq:Diraceq} converting $\gamma^\mu$ to
$\gamma^i$, and then the Dirac equation in these rotating
$(t,x,y,z)$ coordinates with $\gamma^i$ takes the following form,
\begin{equation}
  \Bigl\{ i\gamma^0 \bigl[\partial_t + \Omega(-x\partial_y
  + y\partial_x - \tfrac{i}{2}\sigma^{12})\bigr] + i\gamma^1\partial_x
  +i\gamma^2\partial_y +i\gamma^3\partial_z - m \Bigr\}\psi = 0\;.
\label{eq:Deq}
\end{equation}
In this Dirac equation all the contributions with rotation are 
included in the effective chemical potential $\Omega(-x\partial_y + 
y\partial_x - \tfrac{i}{2}\sigma^{12}) = \Omega\hat J_z$.  This is not 
the case in hydrodynamic approaches; the vorticity is defined with 
derivative, and only the leading order term in the derivative 
expansion are usually picked up.  The solutions of the above Dirac 
equation provide us a complete set of bases.  The positive-energy 
particle solutions with positive and negative helicity take the 
following explicit form in the Dirac representation of $\gamma^i$'s;
\begin{equation}
 u_+ = \frac{e^{-iEt+ip_z z}}{\sqrt{\varepsilon + m}}
  \begin{pmatrix}
    \displaystyle (\varepsilon + m) \phi_{\ell} \\ 
    \displaystyle 0 \\
    \displaystyle p_z\, \phi_{\ell} \\
    \displaystyle ip_{\ell,\,k}\, \varphi_{\ell}
  \end{pmatrix},\;\;
 u_- = \frac{e^{-iEt+ip_z z}}{\sqrt{\varepsilon + m}}
  \begin{pmatrix}
    \displaystyle 0 \\
    \displaystyle (\varepsilon + m) \varphi_{\ell} \\
    \displaystyle -ip_{\ell,\,k}\, \phi_{\ell} \\
    \displaystyle -p_z\, \varphi_{\ell}
  \end{pmatrix},
\label{eq:Dsolu}
\end{equation}
where $\varepsilon \equiv |E+\Omega j|$.  Here $j$ represents the
$z$-component of the total angular momentum and we introduce
$\ell=\ell_+=\ell_--1$ with the azimuthal quantum number $\ell_\pm$
for spin ``up'' and ``down'' states, so that $j=\ell+1/2$ holds for
any spin states.  Also, we defined scalar functions of the radial
momentum as $\phi_\ell=e^{i\ell\theta}J_\ell(p_{\ell,\,k} r)$
and $\varphi_{\ell}=e^{i(\ell+1)\theta}J_{\ell+1}(p_{\ell,\,k} r)$,
which lead to the dispersion relation
$\varepsilon^2 = p_{\ell,\,k}^2+p_z^2+m^2$.  In the same way the
negative-energy antiparticle solutions with positive and negative
helicity are obtained from $v_\pm=i\gamma^2 u_\pm^\ast$ as
\begin{equation}
 v_+ = \frac{e^{iEt-ip_z z}}{\sqrt{\varepsilon + m}}
  \begin{pmatrix}
    \displaystyle -ip_{\ell,\,k}\, \varphi_{\ell}^\ast \\
    \displaystyle -p_z\, \phi_{\ell}^\ast \\
    \displaystyle 0 \\
    \displaystyle (\varepsilon + m) \phi_{\ell}^\ast
  \end{pmatrix},\;\;
 v_- = \frac{e^{iEt-ip_z z}}{\sqrt{\varepsilon + m}}
  \begin{pmatrix}
    \displaystyle -p_z\, \varphi_{\ell}^\ast \\
    \displaystyle -ip_{\ell,\,k}\, \phi_{\ell}^\ast \\
    \displaystyle -(\varepsilon + m) \varphi_{\ell}^\ast \\
    \displaystyle 0
  \end{pmatrix}.
\label{eq:Dsolv}
\end{equation}
As we discuss later, we will compute the vacuum expectation value of
field operators using these basis functions.

Lastly, we mention that our analysis with $u_\pm$ and $v_\pm$ is 
valid for the system with cylindrical symmetry.  In a boundary without 
cylindrical symmetry (e.g. a rotating square), the angular momentum 
$\ell$ is no longer a good quantum number and thus the discussion based 
on the analytic calculation cannot be applied.

\section{Momentum discretization}

In a finite box the momenta should be discrete reflecting the (sharp)
boundary condition imposed on the edge of the box.  We now consider
a cylinder that has a boundary at $r=\sqrt{x^2+y^2}=R$ and is
infinitely long along the $z$-axis.  Then,
$p_z$ is not modified, while the radial momenta should take discrete
values gapped by $\propto R^{-1}$, which was the reason why we denoted
them as $p_{\ell,\,k}$.  Since this discretization property is such
crucial for our quantitative comparisons, let us carefully see how the
discretization condition is physically required.

To this end, we see how the current conservation follows in a
finite-size cylindrical system~\cite{Ambrus:2015lfr}.  For the fermion 
in curved spacetime the vector current conservation law reads,
\begin{equation}
  \nabla_\mu j^{\, \mu} = \frac{1}{\sqrt{|g|}}
  \partial_\mu (\sqrt{|g|}\, j^{\,\mu}) = 0\;,
\end{equation}
where $\nabla_\mu$ represents the covariant derivative and
$j^\mu=\bar{\psi}\gamma^\mu\psi$.  Thus, to keep the total charge
constant in a cylinder, we must impose a condition of no incoming flux
at the spatial boundary as
\begin{equation}
  \int_V dV\, \partial_\alpha (\sqrt{|g|}\;
  \bar{\psi}\gamma^\alpha\psi) \\
  = \int_{\partial V} d\Sigma_\alpha \sqrt{|g|}\;
    \bar{\psi}\gamma^\alpha\psi = 0 \;.
\end{equation}
Here $\alpha$ stands for the spatial components $x,y,z$ in coordinate
space.  In cylindrical coordinates the above condition turns into
\begin{equation}
  R\int_{-\infty}^\infty dz \int_0^{2\pi} d\theta\;
  \bar{\psi}\gamma^r\psi \Bigr|_{r=R} = 0\;.
 \label{eq:boud-cond}
\end{equation}
We note that $\gamma^r\equiv \gamma^1\cos\theta+\gamma^2\sin\theta$
that follows from
$\gamma^1\partial_1 + \gamma^2\partial_2
= \gamma^r\partial_r + r^{-1}\gamma^\theta \partial_\theta$.  For
arbitrary fermionic fields we can expand $\psi(x)$ using the complete
set of $u_\pm(x)$ and $v_\pm(x)$, and then after the
$\theta$-integration which constrains possible combinations of $\ell$,
we find a superposition of four linear independent quantities;
\begin{displaymath}
 \begin{split}
  & J_\ell(p_{\ell-1,k}R) J_\ell(p_{\ell,k'}R)\;,\quad
    J_\ell(p_{\ell,k}R) J_\ell(p_{\ell-1,k'}R)\;,\\
  & J_\ell(p_{\ell-1,k}R) J_\ell(p_{-\ell-1,k'}R)\;,\quad
    J_\ell(p_{\ell,k}R) J_\ell(p_{-\ell,k'}R)\;.
 \end{split}
\end{displaymath}
To realize the fluxless condition for arbitrary $\psi(x)$ we have to
make all of them vanishing and this is possible when the transverse
momenta are discretized as~\cite{Ambrus:2015lfr,Hortacsu:1980kv}
\begin{equation}
  p_{\ell,\,k} = \begin{cases}
     \xi_{\ell,\,k}\;R^{-1}
      \quad\text{for}\quad \ell=0,1,\dots \\
     \xi_{-\ell-1,\,k}\;R^{-1}
      \quad\text{for}\quad \ell=-1,-2,\dots
  \end{cases}
\end{equation}
where $\xi_{\ell,\,k}$ represents the $k$-th zero of $J_\ell(x)$.

The most fundamental quantity to calculate physical observables is
Green's function or the propagator.  The propagator for rotating
systems is modified by the boundary effects at $r=R$ as well as the
non-trivial metric tensor involving $\Omega$.  We can readily
construct the free propagator from $u_\pm(x)$ and $v_\pm(x)$ as
\begin{equation}
  \begin{split}
    S_F^{\alpha\beta}(x,x') & = i \int \frac{dp_0\,dp_z}{(2\pi)^2}
    \frac{1}{2\pi}\sum_{\ell=-\infty}^\infty\sum_{k=1}^\infty
     \frac{2}{[J_{\ell+1}(p_{\ell,k}R)]^2 R^2}\\
    &\qquad \times \frac{e^{-i p^0 (t-t') +ip_z (z-z')}}
    {(p^0+\Omega j)^2-\varepsilon^2+i\epsilon}
    \calS^{\,\alpha\beta}(p;r,\theta,r',\theta')\;.
 \end{split}
\label{eq:prop}
\end{equation}
We should note that the weight in the $\ell$- and $k$-sum are determined 
from the Bessel-Fourier expansion and the following orthogonal relation,
\begin{equation}
  \int_0^R dr\, r\, J_\ell(p_{\ell,k}r)J_\ell(p_{\ell,k'}r)
  = \frac{R^2}{2}\delta_{kk'} [J_{\ell+1}(p_{\ell,k}R)]^2\;,
\end{equation}
and we can numerically verify that the following approximation works
at excellent precision for not too large $\ell$ (for example, for
$\ell\sim 100$ and $k\sim 10$, the deviation is $\sim 1\%$ and the
agreement is better for smaller $\ell$);
\begin{equation}
  \frac{2}{[J_{\ell+1}(p_{\ell,k}R)]^2 R^2} \approx
  p_{\ell,k}\,\Delta p_{\ell,k}\;,
\end{equation}
where $\Delta p_{\ell,\,k} \equiv p_{\ell,\,k+1}-p_{\ell,\,k}$.  
This approximated form is useful to think of the continuum limit with
$R\to\infty$.
Using new notations, $\phi_\ell(r,\theta)\equiv \phi_\ell$, 
$\phi_\ell(r',\theta')\equiv \phi'_\ell$ and so on,
we can parametrize the matrix elements in the propagator as
\begin{equation}
  \mathcal{S}(p;r,\theta,r',\theta') =
    \begin{pmatrix}
      \mathcal{M}_+ & \mathcal{N}_+ \\
      \mathcal{N}_- & \mathcal{M}_-
  \end{pmatrix}\;,
\end{equation}
with
\begin{align}
  \mathcal{M}_\pm &\equiv 
    \begin{pmatrix}
      (\pm p_0 + m) \phi_\ell\phi'_\ell & 0 \\
      0  & (\pm p_0 + m)  \varphi_\ell\varphi'_\ell 
    \end{pmatrix} \\
  \mathcal{N}_\pm &\equiv 
    \begin{pmatrix}
      -p_z \phi_\ell\phi'_\ell 
      & \pm ip_{\ell,k}\phi_\ell\varphi'_\ell \\
      \mp ip_{\ell,k} \varphi_\ell\phi'_\ell 
      & p_z \varphi_\ell\varphi'_\ell
    \end{pmatrix}\;,
\end{align}

\section{Rotating and yet unchanged condensate}

Let us take an explicit example to calculate the field expectation
value in the rotating frame.  An effective model with four-fermion
interaction is an ideal setup for this purpose to investigate the fate
of the chiral condensate.  The effective Lagrangian is
\begin{equation}
  \Lag_\text{4-fermi} = \bar{\psi} \bigl[ i\gamma^\mu
    (\partial_\mu+\Gamma_\mu) - m \bigr] \psi
    + \frac{G}{2}\bigl[ (\bar{\psi}\psi)^2
    + (\bar{\psi}i\gamma^5\psi)^2 \bigr]\;.
\end{equation}
The effective action at the one-loop order in the mean-field
approximation reads,
\begin{equation}
  \Gamma_{\text{eff}}[m(r)] = \int d^4x\,
  \frac{m(r)^2}{2G} - \Tr\ln\bigl[ \partial_\mu+\Gamma_\mu-m(r) \bigr]\;.
\end{equation}
From the condition, $\delta\Gamma_{\text{eff}}[m]/\delta m(r)=0$, 
we can write down the gap equation as
\begin{equation}
  m(r) = G\, \tr\, S_F(x,x)\;.
\end{equation}
Here $S^{\alpha\beta}_F(x,y)$ represents the free fermion propagator 
with mass $m(r)$.  It is technically difficult to solve this functional 
gap equation self-consistently~\cite{Buballa:2014tba}, and in the
present work we will work in the local density
approximation~\cite{Jiang:2016wvv}.  That is, we solve $m(r)$ at each
$r$ as if $m(r)$ were an $r$-independent variable.  We can justify
such an approximate treatment for $\partial_r m\ll m^2$.  Now under
this approximation, we can perform the one-loop integration of the gap
equation as
\begin{equation}
 \begin{split}
 \frac{m(r)}{G}
  = \frac{i}{(2\pi)^2}\int_{-\infty}^\infty dp_z
   &\sum_{\ell=-\infty}^\infty\sum_{k=1}^\infty
   \frac{2}{[J_{\ell+1}(p_{\ell,k}R)]^2 R^2} \\
  &\times 
   \int_{-i\infty+\Omega j}^{i\infty+\Omega j}\frac{dp_0}{2\pi}
   \frac{\tr[\calS(p;r,\theta)]}{p_0^2 - \varepsilon^2}\;.
 \end{split}
\end{equation}
We can explicitly take $\tr[\calS(p,r,\theta)]$ to simplify the
right-hand side.  We note that $\calS(p,r,\theta)$ generally has
the $\theta$-dependence, but its trace does not depend on $\theta$ any
more as seen from
\begin{equation}
  \tr[\calS(p,r,\theta)] = 2m\bigl[ J_\ell(p_{\ell,\,k} r)^2
  + J_{\ell+1}(p_{\ell,\,k} r)^2 \bigr]\;.
\end{equation}
Then, after the $p_0$-integration, the gap equation in the local
density approximation leads to
\begin{equation}
 \begin{split}
  \frac{m}{G} =  \frac{m}{(2\pi)^2}\int_{-\infty}^\infty & dp_z
   \sum_{\ell=-\infty}^\infty\sum_{k=1}^\infty
   \frac{2}{[J_{\ell+1}(p_{\ell,k}R)]^2 R^2}\\
  &\times \frac{J_\ell(p_{\ell,\,k}r)^2 + J_{\ell+1}(p_{\ell,\,k}r)^2}
   {\varepsilon} \theta\,\bigl(\varepsilon - |\Omega j| \bigr)\;.
 \end{split}
\label{eq:gapeqtheta}
\end{equation}
In the same way as the finite-density system, the effect of the
rotation appears only in the form of the theta function constraint
which represents an effective chemical potential of
$|\Omega j|=|\Omega (\ell+1/2)|$ induced by rotation.  Therefore, the
modification caused by rotation comes out from the contribution with
$\varepsilon < |\Omega j|$.  Here we note that $m(r)$ has been assumed 
to be independent of $\Omega$, but we can confirm this self-consistently.

If we make an approximation of $R\sim\infty$ and treat the problem
with a continuous transverse momentum instead of $p_{\ell,\,k}$, the
rotation and the finite chemical potential appear identical in the gap
equation.  In a rotating frame, however, the causality constraint,
$\Omega R\leq 1$, prevents us from taking arbitrarily large $R$.  Once
the boundary at $r=R$ is properly taken into account, there is no such
mode that satisfies $\varepsilon < |\Omega j|$, as we see below.  It
is easy to understand this  from the discretization condition;
$\varepsilon$ becomes minimized at $p_z=m=0$ and $k=1$, so that we can
see, for $\ell\ge 0$,
\begin{align}
  \varepsilon - \Omega|\ell + 1/2|
  &\geq  \frac{1}{R}\Bigl[ \xi_{\ell,\,1} - \Omega R (\ell + 1/2)
  \Bigr] \notag\\
  &\geq \frac{1}{R}\Bigl[ \xi_{\ell,\,1} - (\ell + 1/2)\Bigr] > 0\,,
\label{eq:IRmode}
\end{align}
where we used the causality constraint $\Omega R \leq 1$ and an inequality
known for the zeros of the Bessel function, that
is~\cite{GIORDANO1983221},
\begin{equation}
  \xi_{\ell,\,1} > \ell + 1.855757\ell^{1/3} + 0.5\ell^{-1/3}
  \qquad (\ell\ge 1)\;,
\end{equation}
from which we can show $\xi_{\ell,\,1}>\ell+1/2>0$ for $\ell\ge 1$ and
also we can check $\xi_{\ell,\,1}=2.40483>1/2$ for $\ell=0$.  In
the same way we can also prove that $\varepsilon < |\Omega j|$ is
never realized for $\ell<0$.  We note that a similar discussion is
applicable to bosonic systems; $\varepsilon - |\Omega \ell| >0$ (for
zero-spin bosons).  This ensures that the bosonic thermal distribution
in a rotating frame, $[e^{\beta(\varepsilon - \Omega \ell)} -1 ]^{-1}$
does not exhibit instability (see, for example, discussions in
Ref.~\cite{Davies:1996ks}).

Now we expect that the analogy between density and rotation could help us 
to clarify the above physics.  In finite density systems microscopic 
quantities, such as the Dirac eigenvalue, are affected by chemical potential.  
The density effect on macroscopic quantities at zero temperature can however 
be visible only for the chemical potential lager than the mass threshold; 
this is well-known as the Silver Blaze problem in finite density QCD.  Since 
rotation seems to generate the alignment of the azimuthal angular momentum 
and spin of each rotating fermion, the pairing state with zero total angular 
momentum might no longer be energetically most favored.  Contrary to such an 
intuitive picture, as we have discussed above, the effective chemical 
potential $|\Omega j|$ can never exceed the threshold $p_{\ell,\,1}$.

Even though there is no rotation effect at zero temperature, it is an
intriguing question how $m(r)$ looks like in the local density
approximation with the boundary condition.  To solve the gap
equation~\eqref{eq:gapeqtheta} we need to introduce a ultraviolet (UV)
regulator, which is a part of the four-fermion interacting model
that is non-renormalizable.  We do this by inserting a smooth cutoff
function into the summation as follows;
\begin{equation}
  f(p\,;\Lambda) = \frac{\sinh(\Lambda/\delta\Lambda)}
    {\cosh[\tilde{\varepsilon}(p)/\delta\Lambda]
     + \cosh(\Lambda/\delta\Lambda)}
\label{eq:cutoff}
\end{equation}
with $\tilde{\varepsilon}\equiv\sqrt{p_{\ell,\,k}^2 + p_z^2}$. 
This function is suppressed for $\tilde{\varepsilon}>\Lambda$ and the
suppression smoothness is tuned by a parameter $\delta\Lambda$.  In
the limit of $\delta\Lambda/\Lambda \to 0$ we see that
$f(p\,;\Lambda)$ is reduced to the step function,
$\theta(1-\tilde\varepsilon)=\theta\,(\Lambda^2-p_{\ell,\,k}^2-p_z^2)$.
It is very important to adopt a smooth cutoff because we make a
discrete sum over $p_{\ell,\,k}$ and a sharp cutoff would affect the
sum in a discontinuous way, leading to artificial oscillatory
behavior.  With our choice, as we checked in Ref.~\cite{Chen:2015hfc},
we can perform a systematic analysis on whether our results are robust
and free from cutoff artifact.

\begin{figure}
 \begin{center}
   \includegraphics[width=1\columnwidth]{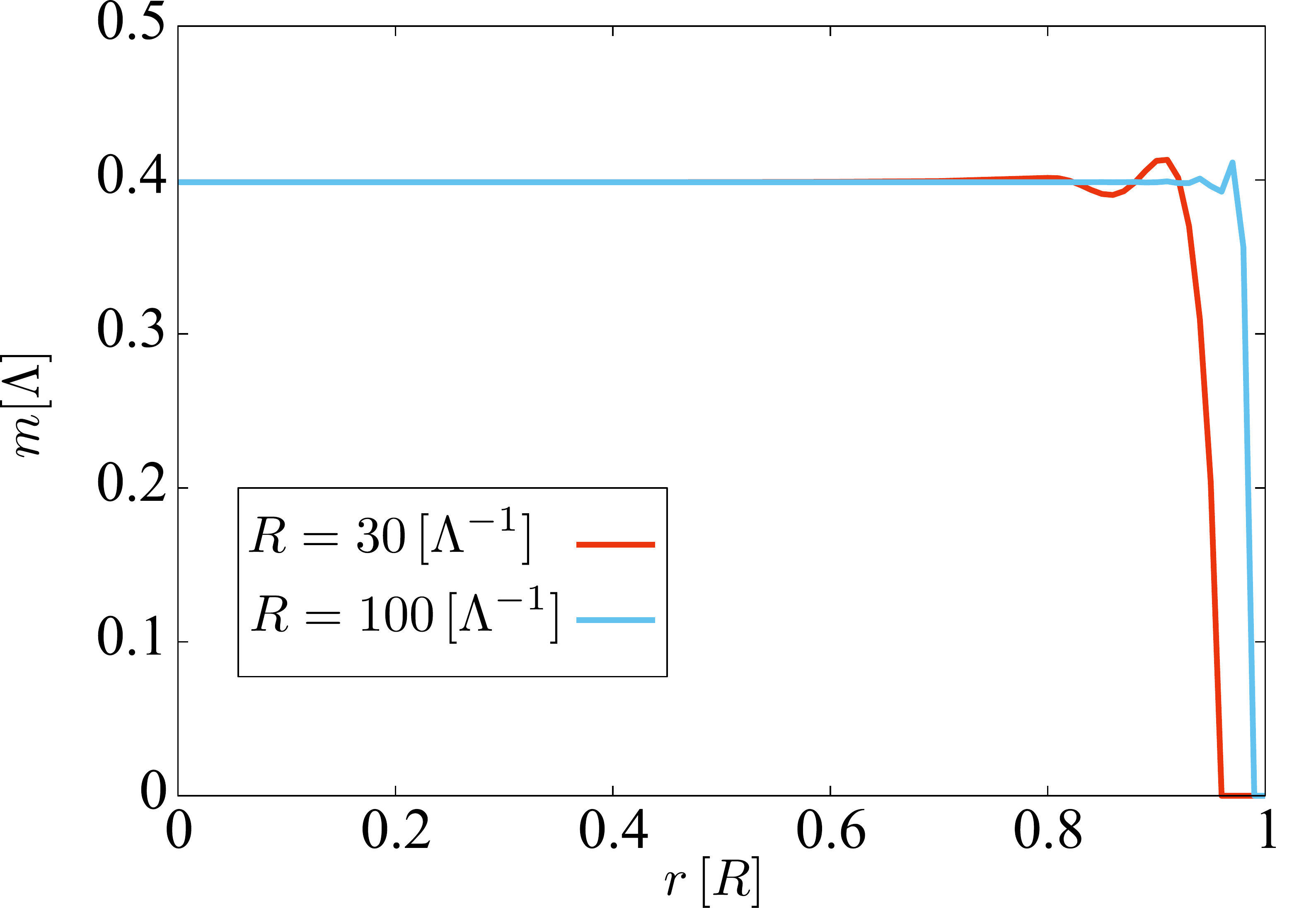}
    \caption{Inhomogeneous dynamical mass as a function of the radial
    coordinate $r$.  Apart from the very vicinity of the boundary
    $\sim R$, the position dependence is mild enough to justify the
    local density approximation and Eq.~\eqref{eq:gapeqtheta}.  The
    vanishing mass at the boundary is a consequence from the
    condition~\eqref{eq:boud-cond}. As $R$ increases, the
    oscillation behavior becomes small, and eventually vanishes.}
    \label{fig:mass}
  \end{center}
\end{figure}

We numerically solved the gap equation~\eqref{eq:gapeqtheta} with
$f(p;\Lambda)$ inserted, with the following parameters:
\begin{align}
  R &= 30\,[\Lambda^{-1}]\,,100\,[\Lambda^{-1}]\,,
   \quad \delta\Lambda = 0.05\, [\Lambda]\,. \notag \\
  G &= 12\,[\Lambda^{-2}] = 0.61G_c\,,
   \quad G_c = 19.65\,[\Lambda^{-2}]\,.
  \label{eq:parameters}
\end{align}
where $G_c$ denotes the critical coupling calculated with 
Eq.~\eqref{eq:cutoff}, $R\to\infty$ and $\delta\Lambda/\Lambda = 0.05$%
~\cite{Chen:2015hfc}.  Here for $\Lambda \simeq 1\,\text{GeV}$ that is 
the common choice in four-fermion models used for the strong interaction 
physics, the system size of the above choice corresponds to the typical 
radius scale of the heavy ion, namely, $R = 30\Lambda^{-1} \sim 6\,
\text{fm}$.  Figure~\ref{fig:mass} is a plot to show the $r$ dependence 
of the dynamical mass.  We can confirm that the local density approximation
is self-consistently reliable unless we go to the very vicinity of the
boundary $\sim R$ where $|\partial_r m / m^2|\gg 1$ is no longer the
case.  Also in Fig.~\ref{fig:mass} we see an oscillational behavior.
Such an oscillation is the cutoff artifact, which vanishes in the 
continuum limit. Indeed as $R$ increases the oscillation point comes 
closer to $r=R$. It is clear that the spatial inhomogeneity of $m$ is 
eventually washed out in the limit of $R\to\infty$. 
Contrary to this, the boundary effect is generally enhanced for small 
$R$, as shown in Fig.~\ref{fig:mass}.  At the same time, the cutoff 
artifact in the $\ell$- and $k$-sum becomes larger (more badly 
oscillating) because the spacing in discrete $p_{\ell,\,k}$ grows as 
$R$ decreases. Furthermore although the magnitude of the dynamical mass 
is quite sensitive to the coupling $G$, the boundary effect is 
irrelevant to the coupling. From numerical calculation, we have 
actually confirmed that the structures of the spatial profile, i.e., 
both the plateau at $0\leq r\lesssim 0.8R$ and the oscillational 
behavior at $r\gtrsim 0.8R$ are unchanged even if $G$ is changed.

\section{Anomalous coupling to the rotation}

So far we have seen that the rotation does not affect the condensate.
(Microscopic quantities, e.g., the Dirac eigenvalue can however be 
affected by rotation, in analogy to the finite density effect.)
Nevertheless,the rotation can change the physical properties via 
anomalous coupling.  Here let us pick up two well-known such examples.

The first example is the finite temperature.  We know that the
rotating fermionic system develops an axial current at finite
$T$~\cite{Vilenkin:1979ui,Vilenkin:1980zv} as
\begin{equation}
  \bj_{\text{A}}^{\,\text{CVE}} = \frac{T^2\bOmega}{12}
  + \frac{\Omega^2\bOmega}{48\pi^2} + O(\Omega R^{-2}) \quad \text{for}\ 
  T\gg R^{-1} \geq \Omega
 \,,
  \label{eq:CVE}
\end{equation}
where we drop the finite size corrections of $O(\Omega R^{-2})$.
Intuitively we can understand this in the following manner.  For
sufficiently high $T\gg R^{-1}$, it is unlikely that the bulk
constraint at the boundary remains relevant because of the thermal
screening and we can safely neglect the boundary effects.  Indeed, in
this case of $T\gg R^{-1}$, the first term should be much larger than
the second one because the causality constraint demands
$R^{-1} \geq \Omega$ and thus $T\gg\Omega$.  Interestingly, it is known
that this coefficient of the term $\propto T^2$ is related to the
chiral anomaly coming from not gauge fields but Riemann tensors.  We
can therefore say that the rotation effect becomes visible thanks to
the coupling to the gravitational chiral
anomaly~\cite{Landsteiner:2011cp,Basar:2013qia}.  (This terminology
might be a little confusing; the genuine gravity is irrelevant and
what does matter is the chiral anomaly coupled to the Riemann and
stress tensors.)

It would be a very interesting question whether the current still
persists for $T\lesssim R^{-1}$, and to clarify this, we should do the
microscopic calculation with the boundary effects, which was already
pointed out in Ref.~\cite{Vilenkin:1980zv}.  Then, the only change
from the zero to the finite temperature results is how the
effective chemical potential appears, i.e.\
$\theta\,(\varepsilon-|\Omega j|)$ should be replaced with the Fermi
distribution function $[1+e^{-(\varepsilon-|\Omega j|)/T}]^{-1}$,
and then we see that $\Omega$ dependence remains even for
$\varepsilon>|\Omega j|$.  Such an explicit calculation of the chiral
vortical effect in a finite size system will be reported
elsewhere.

Second, we shall turn to magnetized rotating matter as discussed in
Ref.~\cite{Chen:2015hfc}.  Under a strong magnetic field, the Landau
wave-function is localized and can be even more squeezed than the
system size if $\sqrt{eB} \gg R^{-1}$.  Then, the boundary effects are
essentially irrelevant.  Also, the energy dispersion relation of
fermions with $B$ is Landau-quantized and the dynamics of the
magnetized fermions is dominated by the Landau zero mode, which is
independent of the angular momentum.  This is quite different from
rotating fermions without $B$ for which the IR modes are gapped as
seen in Eq.~\eqref{eq:IRmode}.  Therefore, there always exist
low-energy modes that are Pauli blocked, and thus, with help of finite
$B$, the rotation comes to affect the system even at zero
temperature.  This is an hand-waving explanation for the reason why it
has been observed in Ref.~\cite{Chen:2015hfc} that the rotation
affects the chiral condensate.

Interestingly, in this case too, the quantum anomaly plays a crucial
role.  Unlike the temperature for which the gravitational mixed
anomaly is relevant, the well-known standard chiral anomaly in terms
of the gauge field is sufficient to understand how the rotation and
the magnetic field can induce a finite density.  To see this
explicitly, let us consider a Dirac fermion in the magnetic field
$\bB=B\hat{\bz}$ \textit{without} rotation.  The Lagrangian density is
simply $\Lag = \bar{\psi} i\gamma^i (\partial_i + ie A_i)\psi$,
where $A_i =(0,By/2,-Bx/2,0)$ in the symmetric gauge choice.  Now,
we shall perform the ``Floquet transformation''~\cite{Bukov:2014} or
go to the rotating frame by changing,
\begin{equation}
 \psi \to \exp(\gamma^1 \gamma^2 \Omega t/2)\, \psi\;,
\label{eq:floquet}
\end{equation}
together with the coordinate transformation by
$x\to (\cos\Omega t)x-(\sin\Omega t)y$ and
$y\to (\cos\Omega t)y+(\sin\Omega t)y$. 
Then, the Lagrangian density after the transformations reads,
\begin{equation}
 \begin{split}
  \Lag = \bar{\psi} [ i\gamma^0 \partial_t
  +i\gamma^1( \partial_x + ieBy/2 ) & + i\gamma^2 ( \partial_y -ieBx/2 ) \\
  & +i\gamma^3 \partial_z + (\Omega/2) \gamma^3 \gamma_5] \psi\,.
 \end{split}
\label{eq:rotated_Lag}
\end{equation}
Here, we can regard the last term proportional to $\Omega/2$ as an
axial gauge field or the chiral shift~\cite{Gorbar:2009bm}, and as
calculated in Ref.~\cite{Ebihara:2015aca}, a finite density is induced
from the quantum anomaly coupled with the chiral shift term and the
magnetic field as
\begin{equation}
 n_\text{spin} = \frac{eB\Omega}{4\pi^2}\,,
\label{eq:cpe} 
\end{equation} 
which explains the expression for the density obtained in
Ref.~\cite{Hattori:2016njk}.

In the above discussions one might have realized that
Eq.~\eqref{eq:rotated_Lag} is not really the Lagrangian density with
$B$ in a rotating frame, in which more terms like
$\Omega(-x\partial_y+y\partial_x)$ should appear. These terms do not enter 
Eq.~\eqref{eq:rotated_Lag} because the Floquet transformation 
Eq.~\eqref{eq:floquet} does not accompany the rotation of the orbital part. 
In fact we can show that the above anomalous density picks up a contribution 
from the spin part only.

Because we already know the complete expression for the thermodynamics
potential or the free energy with both $B$ and $\Omega$ in
Ref.~\cite{Chen:2015hfc}, it is easy to take its chemical potential
derivative and compute the density.  The free energy under strong $B$
enough to discard the boundary effects reads,
\begin{equation}
 \begin{split}
  F &= -\frac{1}{\pi R^2}
  \sum_{q=\pm} \int_{-\infty}^\infty \frac{dp_z}{2\pi} 
  \sum_{n=0}^\infty \alpha_n \\
  &\qquad\times \sum_{\ell=-n}^{N-n} 
  \biggl\{ \frac{\varepsilon + q\Omega j + q\mu}{2}
  +T \ln \bigl[1+e^{-(\varepsilon + q\Omega j + q\mu)/T}\bigr] \biggr\},
 \end{split}
 \label{eq:F}
\end{equation}
where $\alpha_n=2-\delta_{n,0}$, $j=\ell+1/2$, $N=eBR^2/2$, and
$\varepsilon = \sqrt{p_z^2+2neB}$.  By differentiating $F$ with respect
to $\mu$ and taking the $T\to0$ limit, the number density in the lowest
Landau approximation turns out to be
\begin{equation}
  n_\text{total} = \frac{\Omega}{\pi^2 R^2} \sum_{\ell=0}^{N}(\ell+1/2)
  = \frac{eB\Omega}{4\pi^2}(N+1)\;.
\label{eq:denrot}
\end{equation}
We see that, in addition to the anomaly-induced density in
Eq.~\eqref{eq:cpe}, we have an extra contribution from the orbital
angular momentum $\ell$, which makes a contrast to the result in
Ref.~\cite{Hattori:2016njk}. We emphasize that the total angular
momentum (i.e., both the orbital and spin angular momentum) contribute
this anomalous effect. 
What we can learn from the above exercises is that the rotation can affect 
the thermodynamic properties and thus modify the condensate if a strong 
magnetic field is imposed.

We already mentioned that the intermediate region is difficult to
investigate.  For the temperature effect, what happens for
$T\lesssim R^{-1}$ still needs careful considerations, and in the same
way for the magnetic effect, it would be a quantitatively subtle
question to study the regime for $\sqrt{eB}\lesssim R^{-1}$.  In most
of physics problems involving quarks and gluons, either
$T\gg R^{-1}$ (in a quark-gluon plasma) or $\sqrt{eB}\gg R^{-1}$ (in a
neutron star) would be realized, but for future applications to
table-top experiments, a more complete treatment over the whole regime
would become important.

\section*{Acknowledgements}
The authors thank Xu-Guang Huang, Koich Hattori, and Yi Yin
for useful discussions.  S.~E.\ also thanks Leda Bucciantini, Yosh\textsl{•}imasa 
Hidaka, and Takashi Oka for discussions.  K.~F.\ is grateful for a warm 
hospitality at Institut f\"{u}r Theoretische Physik,  Universit\"{a}t 
Heidelberg, where K.~F.\ stayed as a visiting  professor of EMMI-ExtreMe 
Matter Institute/GSI and a part of this work was completed there.
This work was supported by Japan Society for the Promotion of Science 
(JSPS) KAKENHI Grant Nos.\ 15H03652 and 15K13479 (K.~F.), 
Grant-in-Aid for JSPS Fellows Grant No. 15J05165 (K.M.).

\bibliographystyle{h-elsevier}
\bibliography{rotation}

\end{document}